\documentclass[structabstract]{aa} 
\usepackage{graphicx}
\usepackage{natbib}

\usepackage{amsmath}
\usepackage{amssymb}
\usepackage{subeqnarray}
\usepackage{layouts}
\usepackage{url}

\usepackage{color}
    \definecolor{Blue}{rgb}{0.0,0.0,1.0}
    \definecolor{Red}{rgb}{1.0,0.0,0.0}
    \definecolor{Green}{rgb}{0.0,1.0,0.0}

\newcommand{\be}{\begin{equation}}                                 
\newcommand{\ee}{\end{equation}}                                   
\newcommand{\bea}{\begin{eqnarray}}                                
\newcommand{\eea}{\end{eqnarray}}                                  
\newcommand{\nn}{\nonumber}                                                   
\definecolor{gray}{rgb}{.6,.6,.6}                                  %
\definecolor{green}{rgb}{0,.6,0}                                   %
\definecolor{red}{rgb}{0.6,0,0}                                    %

\begin{document}
\title{Towards modeling quasi-periodic oscillations of microquasars with oscillating slender tori}
%
\author{     G. P. Mazur\inst{1, 3}
\and           F. H. Vincent\inst{1}
\and           M. Johansson\inst{4}
\and           E. {\v S}ramkov{\'a}\inst{5}
\and           G. T{\"o}r{\"o}k\inst{5}
\and           P. Bakala\inst{5}
\and           M.~A.~Abramowicz\inst{1, 2, 5}
}
\institute{
              N. Copernicus Astronomical Centre
	   \\ \email{fvincent@camk.edu.pl}
\and          Physics Department, Gothenburg University,
               SE-412-96 G{\"o}teborg, Sweden
                 \\ \email{marek.abramowicz@physics.gu.se}
\and         Physics Department, Warsaw University,
                 \\ \email{gmazur@camk.edu.pl}
\and         Chalmers University of Technology
                  \\ \email{matzjb@yahoo.se}
\and         Physics Department, Silesian University
                  \\ \email{terek@volny.cz}
                  \\ \email{pavel.bakala@fpf.slu.cz}
                  \\ \email{sram\_eva@centrum.cz}
}
   \date{Received ; accepted }
\abstract 
   {One of the often discussed models for high-frequency quasi-periodic oscillations
   of X-ray binaries is the oscillating torus model, which considers 
   oscillation modes of slender accretion tori.}
   {Here, we aim at developing this model by considering the observable signature
   of an optically thick slender accretion torus subject to simple periodic deformations.}
   {We compute light curves and power spectra of a slender accretion torus
   subject to simple periodic deformations: vertical or radial translation, rotation, expansion, and shear.}
   {We show that different types of deformations lead to
\textit{very different} Fourier power spectra and therefore could be
observationally distinguished.}%
   {This work is a first step in a longer term study of the observable
   characteristics of the oscillating torus model. It gives promising perspectives
   on the possibility of constraining this model by studying the observed power spectra
   of quasi-periodic oscillations.}

   \keywords{Accretion, accretion disks -- Black hole physics -- Relativistic processes
               }
\authorrunning{Mazur et al.}\titlerunning{Oscillating tori for modeling QPOs}
\maketitle



\section{Introduction}

Some microquasars exhibit high-frequency (millisecond) quasi-periodic oscillations 
(QPOs), which in this work always refer to high-frequency oscillations; low-frequency
QPOs are not considered here.
These high-frequency oscillations are characterized by a narrow peak in the source power 
spectrum~\citep[see][for a review]{vanderklis04,remillard06}. 
As the timescale of this variability is of the same order as the Keplerian orbital period
at the innermost stable circular orbit of the central black hole, it is highly probable
that these phenomena are linked with strong-field general relativistic effects. {The other different low-mass X-ray binaries that contain neutron stars display QPOs on these prominent
timescales as well. There is no consensus on the possible QPO origin and uniformity across different sources yet.}

A variety of models have been developed so far to account for {the} high-frequency
variability. 
{\citet{ste-vie:1999} and \citet{ste-etal:1999} suggest that QPOs could arise from the modulation of the X-ray flux by the periastron precession and the Keplerian frequency of blobs of matter orbiting in
an accretion disk around the central compact object.}
Also QPOs could be due to modulation of the X-ray flux by oscillations 
of a thin accretion disk surrounding the 
central compact object~\citep[see][]{wagoner99,kato01}.
\citet{fragile01} propose that QPOs are due to the modulation
of the X-ray flux caused by a warped accretion disk surrounding
the central compact object.
Pointing out the 3:2 ratio of some QPOs in different sources,
\citet{abramowicz01} have proposed a resonance model
in which these pairs of QPOs are due to the beat between the
Keplerian and epicyclic frequencies of a particle orbiting
around the central compact object.

{General relativistic ray tracing of radiating or irradiated hot spots orbiting around black holes and neutron stars was founded at the turn of the millennium. For instance, \cite{karas:1999} discusses already sophisticated numeric modeling of observable modulation from clumps distributed around certain preferred circular orbits. Later, \citet{schnittman04} investigated in great detail predictions of a model of hot spot radiating isotropically on nearly circular equatorial orbits. In their study, the hypothetic resonance between the Keplerian and radial epicyclic frequencies gives rise to peaks in the modeled power spectrum.}
\citet{tagger06} advocate the fact that QPOs in microquasars are
due to the triggering of a Rossby wave instability in the accretion
disk surrounding the central compact object. Ray-traced light curves
have been recently developed for this model by~\citet{vincent12}. 
Finally, models involving oscillations of accretion tori have been
developed. As our work is a follow-up of these past studies,
we present them in more details.

The first study of a QPO model involving thick accretion structure (tori)
was developed by~\citet{rezzolla03}, who showed that p-mode oscillation
of a numerically computed accretion torus can generate QPOs. Ray-traced light curves and
power spectra of this model were derived by~\citet{schnittman06}. 
The analysis of analytical accretion tori (for which all physical quantities are
known analytically throughout the structure and at all times) as a model for
QPOs was initiated by~\citet{bursa04}, who performed simulations of
ray-traced light curves and power spectra of an optically thin oscillating slender torus. 
This model took only into account simple vertical and radial sinusoidal motion
of a circular cross-section torus. 
To allow a more general treatment, a series of theoretical works were dedicated to
developing a proper model of general oscillation modes of a slender
or non-slender perfect-fluid hydrodynamical accretion 
torus~\citep{abramowicz06,blaes06,straub09}. Considering hydrodynamical
perfect-fluid tori is interesting in terms of simplicity as all computations
can be derived analytically and all physical quantities are known analytically
at all times.

The aim of this article is to strengthen the perfect-fluid
oscillating slender torus model for QPOs by progressing towards
determining its observable signature. 
Here, we present a first step towards this goal, which consists
of determining the observable signature of simple deformations of
a slender torus in the Schwarzschild metric. 
We do not use the mathematically well-defined
oscillation formalism derived by~\citet{blaes06}. We rather consider
the impact on the observables of elementary deformations of
the torus cross-section: translation, rotation, expansion, and shear.
We simulate light curves and power spectra of such
tori subject to these simple deformations, taking into account all relativistic
effects that will affect radiation by using a ray-tracing code.
{The focus of this article is thus to determine the impact of the geometrical
change of a slender torus on the observed flux variation. We do not consider
the physics of the matter forming the torus.}

This model is extremely simple and does not
claim to give a realistic view of the nature of QPOs. However, 
it does allow determination in the simplest possible framework
of whether different kinds of motions of a toroidal accretion structure
lead to potentially observable differences in the light curves.
This work thus makes it possible to go one step further in the development
of the oscillating torus model for QPOs.


Section~\ref{sec:equil} describes the equilibrium slender torus, while Section~\ref{sec:oscil}
describes its deformations in terms of translation, rotation, expansion, and shear. 
Section~\ref{sec:simu} shows the ray-traced light curves and 
power spectra of the deformed torus, and Section~\ref{sec:conclu} gives conclusions and 
prospects.


\section{Equilibrium of a slender accretion torus}
\label{sec:equil}

This section derives the equations that describe the accretion torus at equilibrium. 
The spacetime is described by the Schwarzschild metric in the Schwarzschild coordinates $(t, r, \theta, \varphi)$, with geometrical units $c = 1 = G$ and signature $(-, +, +, +)$. The line element has then the following form:
\bea
  \mathrm{d}s^2 &=&
  g_{tt} \mathrm{d}t^{2} 
  	+ g_{rr}\mathrm{d}r^2 + g_{\theta\theta}\mathrm{d}\theta^{2} 
		+ g_{\varphi\varphi}\mathrm{d}\varphi^2\nn \\
  &=&
  -\left(1-\frac{2 M }{r}\right) \,\mathrm{d}t^2 
  + \left(1-\frac{2 M }{r}\right)^{-1}\, \mathrm{d}r^2  \nn \\
  &&
  + \,r^2\left(\mathrm{d}\theta^2+\mathrm{sin}^2 \theta \mathrm{d}\varphi^2\right),
\label{eq:schmet}
\index{Sch metric}
\eea
where $M$ is the black hole mass. In the remaining article,
we use units where this mass is $1$.

In this spacetime, we consider an axisymmetric, non-self-gravitating, perfect fluid, constant
specific angular momentum, which circularly orbits the accretion torus. This torus is assumed to be
slender, meaning that its cross section is small as compared to its central radius.

\subsection{Fluid motion}

As the fluid follows circular geodesics, its 4-velocity can be written as
\be
u^{\mu} = A\left( \eta^{\mu}+\Omega \xi^{\mu}\right),
\label{eq-4vel}
\ee
where $\Omega$ is the fluid's angular velocity, and $\eta^{\mu}$ and $\xi^{\mu}$ are the Killing
vectors associated with stationarity and axisymmetry,
respectively. The constant $A$ is given by imposing the normalization
of the 4-velocity, $u^{\mu}u_{\mu}=-1$.

As spacetime is stationary and axisymmetric, the specific energy $\mathcal{E}$
and specific angular momentum $\mathcal{L}$ are defined as
\bea
\mathcal{E} &=& -\eta^{\mu}u_{\mu} = -u_t  \\ \nn
\mathcal{L} &=& \xi^{\mu}u_{\mu} = u_\varphi, 
\eea
which are geodesic constants of motion.

The rescaled specific angular momentum $\ell$ is defined according to
\be
\ell \equiv \frac{\mathcal{L}}{\mathcal{E}} = -\frac{u_{\varphi}}{u_t}.
\ee

We assume this rescaled specific angular momentum
to be constant throughout the torus:
\be
\ell = \ell_0 = {\rm const}.
\ee

The 4-acceleration along a given circular geodesic followed by the fluid is
\be
a_{\mu} = u^{\nu}\nabla_{\nu}u_{\mu} = -\frac{1}{2\mathcal{U}} \partial_{\mu}\mathcal{U},
\ee
where $\mathcal{U}$ is the effective potential~\citep[see, e.g.,][]{abramowicz06}:

\be
\mathcal{U} = g^{tt} + \ell_0^{2}g^{\varphi\varphi}.
\ee

The radial and vertical epicyclic frequencies for circular motion are related to 
the second derivatives of this potential:
\begin{eqnarray}
\omega^2_r &=& \frac{1}{2}\frac{{\cal
E}^2}{A^2\,g_{rr}}\left(\frac{\partial^2{\cal U}}{\partial
r^2}\right), \nn \\
\omega^2_{\theta} &=& \frac{1}{2}\frac{{\cal
E}^2}{A^2\,g_{\theta\theta}}\left(\frac{\partial^2{\cal
U}}{\partial \theta^2}\right).
\label{epicyclic-requencies-second-derivatives}
\end{eqnarray}

In the Schwarzchild metric, these epicyclic frequencies at the torus center are
\begin{eqnarray}
\omega^2_{r0} &=& \Omega_K^2 \left(1-\frac{6}{r_0}\right), \nn \\
\omega^2_{\theta 0} &=& \Omega_K^2
\label{epifreq}
\end{eqnarray}
where a subscript $0$ denotes here and in the rest of this article
a quantity evaluated at the torus center. The Keplerian angular velocity
is well known: $\Omega_K = r_0^{-3/2}$.

In the rest of this article, the torus central radius will be fixed to the
value $r_0$, satisfying the following condition

\be
\frac{\omega_{\theta 0}}{\omega_{r 0}}=\frac{3}{2}.
\ee

This choice is linked to our goal of applying the deformed
torus model to twin-peak QPOs. It results in

\be
r_0 = 10.8.
\ee

%
\subsection{Surface function}
%


{Following~\citet{abramowicz06}, we define the surface of the torus by
the locus of the zeros of a surface function $f(r, \theta)$.}
Introducing the following set of coordinates
\begin{equation}
x = \left( \sqrt{g_{rr}} \right)_0 \left( \frac{r -
r_0}{r_0}\right),~~
y = \left( \sqrt{g_{\theta\theta}} \right)_0 \left( \frac{\pi/2 -
\theta}{r_0}\right),
\label{small-coordinates}
\end{equation}
\citet{abramowicz06} show that the surface function
is expressed according to
\begin{equation}
f = 1 - \frac{1}{\beta^2}\left(  {\bar \omega}^2_{r0}\, x^2 +
{\bar \omega}^2_{\theta0}\, y^2 \right),
\label{function-f-simple-form}
\end{equation}
where
\begin{equation}
{\bar \omega}_{r0} = \frac{\omega_{r0}}{\Omega_K},\quad
{\bar \omega}_{\theta 0} = \frac{\omega_{\theta 0}}{\Omega_K}.
\label{dimensinless-epicycli}
\end{equation}
the parameter $\beta$ 
%
is related to the torus
thickness, and the torus being slender:

\be
\label{eq:beta}
\beta \ll 1.
\ee

\citet{abramowicz06} also show that the $x$ and $y$ coordinates are of order $\beta$.
We thus define a new set of order-unity coordinates:

\begin{equation}
{\bar x} = \frac{x}{\beta}, \quad {\bar y} = \frac{y}{\beta},
\label{magnified-coordinates}
\end{equation}
and:
\begin{equation}
f = 1 - \left(  {\bar
\omega}^2_{r0}\, {\bar x}^2 + {\bar \omega}^2_{\theta0}\, {\bar
y}^2 \right).
\label{function-f-simple-form-2}
\end{equation}

The equilibrium slender torus has therefore an elliptical
cross section. 




\section{Deformations of a slender accretion torus}
%
\label{sec:oscil}

This section is devoted to determining the surface function 
of the torus subject to simple deformations: translation, rotation, expansion, and shear. 
The 4-velocity of the
perturbed fluid is also given.

%
%
\subsection{Deforming the torus}
%

In this section, we consider various simple time-periodic deformations
of the torus cross-section. Our aim is to determine
the new torus surface function corresponding to these deformed states.

All deformations boil down to performing transformations
on the $(\bar{x},\bar{y})$ coordinates of the form:

\bea
\label{coordtrans2}
\bar{x}(t) &=& a_1(t) \bar{x}(0) +a_2(t) \bar{y}(0) +a_3(t), \\ \nn
\bar{y}(t) &=& b_1(t) \bar{x}(0) +b_2(t) \bar{y}(0) +b_3(t), \\ \nn
\eea
where the $a_i$ and $b_i$ functions are simple trigonometric
functions of time. {Table~\ref{tabdefor} gives these functions for
all deformations considered in this article
as a function of a free parameter $\lambda$ 
and of the deformation's pulsation $\omega$. This
pulsation will have the same value for all the simulations
performed in this article. We choose $\omega = \Omega_K$.
We note that this value is a choice as the deformations
we consider are not physically justified, but imposed
for the simplicity of the model. We make here the simplest
and most natural choice of considering the Keplerian pulsation.}

\begin{table*}
\begin{center}
\begin{tabular}{lllllll}
\hline
Deformation            & $a_1$ & $a_2$ & $a_3$ & $b_1$ & $b_2$ & $b_3$ \\
\hline
Radial translation    	& $1$     & $0$ &   $\lambda\,\mathrm{sin}(\omega t)$ & $0$ & $1$ & $0$  \\
Vertical translation     & $1$ & $0$ & $0$ & $0$ & $1$  & $\lambda\,\mathrm{sin}(\omega t)$  \\
Rotation     		  &  $\mathrm{cos}(\omega t)$ &  $\mathrm{sin}(\omega t)$ & $0$ &  $-\mathrm{sin}(\omega t)$ & $\mathrm{cos}(\omega t)$ & $0$ \\
Expansion    		  & $1+\lambda\,\mathrm{sin}(\omega t)$ & $0$ & $0$ & $0$ &  $1+\lambda\,\mathrm{sin}(\omega t)$ & $0$ \\
Radial simple shear   & $1$ & $\lambda\,\mathrm{sin}(\omega t)$ & $0$ & $0$ & $1$ & $0$  \\
Vertical simple shear   & $1$ & $0$ & $0$ &  $\lambda\,\mathrm{sin}(\omega t)$ & $1$ & $0$  \\
Pure shear    		   & $1+\lambda\,\mathrm{sin}(\omega t)$ & $0$ & $0$ &  $0$ & $\left[1+\lambda\,\mathrm{sin}(\omega t)\right]^{-1}$ & $0$  \\
\hline
\end{tabular}
\caption{\label{tabdefor} {Functions $a_i$ and $b_i$ of Eq.~\ref{coordtrans2} for all periodic deformations, $\lambda$ is a free 
	parameter and $\omega$ is the deformation's pulsation.}}
\end{center}
\end{table*}

%
%
%
%
%
%
%

At this stage, the deformed torus we are considering has thus
only two degrees of freedom:

\begin{itemize}
\item[-] the torus thickness $\beta$ parameter, with $\beta \ll 1$,
\item[-] the deformation parameter $\lambda$, with typically $\lambda \approx 1$.
\end{itemize}

The actual values of $\beta$ and $\lambda$ will be chosen
to satisfy the slender torus approximation for all times.

%
%
\subsection{Perturbed fluid 4-velocity}
%

The fluid 4-velocity in the deformed torus is given by
\begin{eqnarray}
u^{\mu} &=& u^{\mu}_0 + \delta u^{\mu},\\ \nn
u^{\mu}_0 &=& A_0(\eta^{\mu} + \Omega_K \xi^{\mu})\\ \nn
\label{four-velocity-general}
\end{eqnarray}
where $u_0^{\mu}$ is the equilibrium 4-velocity already defined
in Eq.~(\ref{eq-4vel}), which is assumed to be everywhere
equal to its central value, in the slender torus limit. 

The value of the perturbation $\delta u^{\mu}$ induced
by the deformation is easily computed by using the coordinate
transformations given in the previous Section. By writing a first-order
expansion of $\bar{x}(t+\delta t)$ and $\bar{y}(t+\delta t)$ for a
small time increment $\delta t$, it is straightforward to obtain the
expressions of $\mathrm{d}\bar{x} / \mathrm{d}t$ and $\mathrm{d}\bar{y} / \mathrm{d}t$,
which are themselves linearly related to $\mathrm{d}r / \mathrm{d}t$
and $\mathrm{d}\theta / \mathrm{d}t$.

An example of this is the case
of radial translation of the torus cross-section:

\bea
\label{eq:pert}
u_r &=& \delta u_r \\ \nonumber
u_\theta &=&\delta u_\theta = 0,
\eea
with
\be
\delta u_r = \frac{\mathrm{d}r}{\mathrm{d}t}u^t=\frac{\lambda \,\omega\, r_0}{ \left( \sqrt{g_{rr}} \right)_0}\mathrm{cos}(\omega t)u^t
\ee
from Table~\ref{tabdefor}.

As the torus is assumed to have a constant rescaled angular momentum $\ell_0$,
the following relation holds between the non-equilibrium temporal and azimuthal
covariant components of the 4-velocity:

\be
u_\varphi = -\ell_0 u_t.
\label{eq:up-ut}
\ee

Adding the normalization condition $u^{\mu}u_\mu=-1$, Eqs.~(\ref{eq:pert})
to~(\ref{eq:up-ut}) fully determines the fluid 4-velocity.
Equivalent formula can be derived for all other deformations.

%



\section{Light curves and power spectra of the deformed torus}
\label{sec:simu}

%
%
\subsection{Ray tracing on the deformed torus}
%
%

Light curves and power spectra of the deformed torus are computed
by using the general relativistic ray-tracing code \texttt{GYOTO}~\citep{vincent11}.
Null geodesics are integrated backward in time from a distant observer's screen 
to the optically thick torus. The zero of the surface function is found numerically 
along the integrated geodesic and the fluid 4-velocity is determined by using 
Eqs.~(\ref{four-velocity-general}). The specific intensity emitted by the undeformed
torus is chosen to be constant throughout the surface. An image is then defined
as a map of specific intensity.

\begin{figure*}[htbp]
\begin{center}
\includegraphics[width=0.75\hsize]{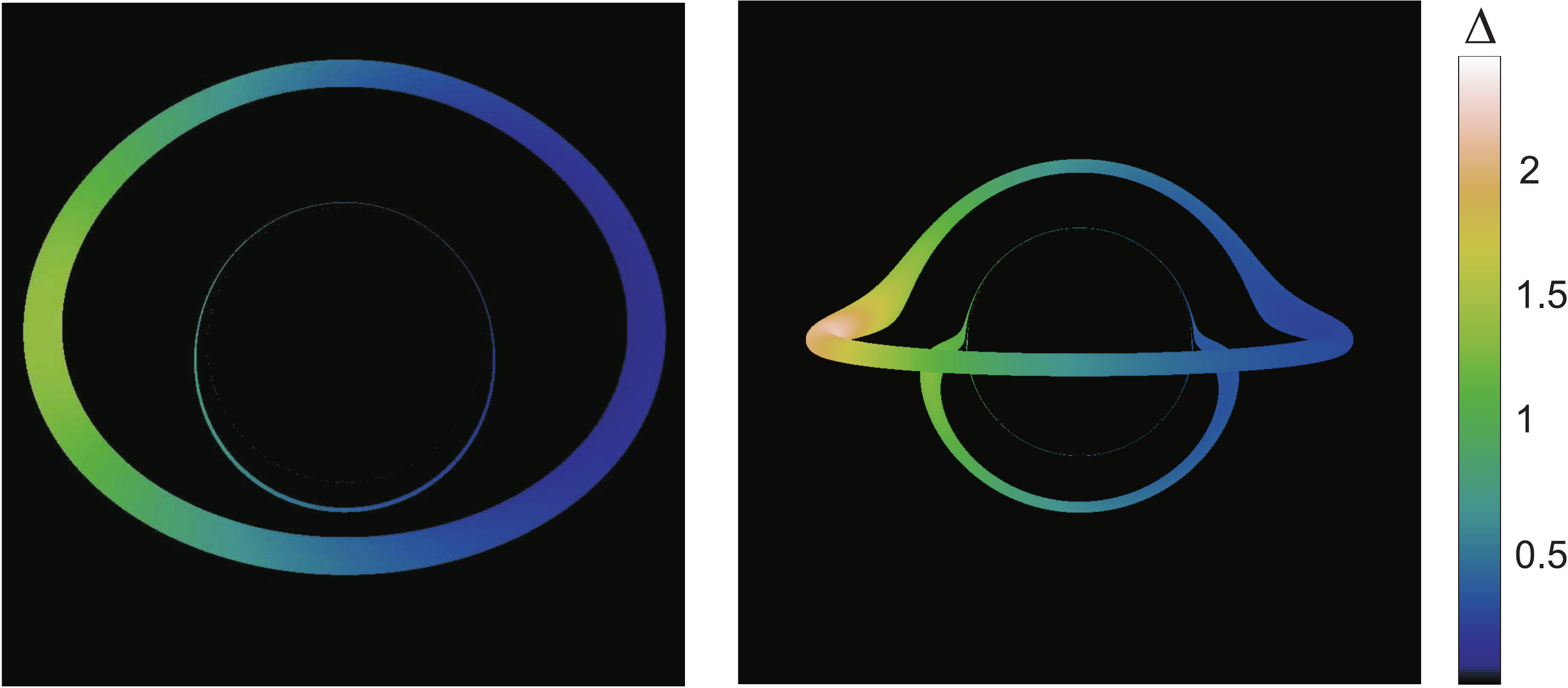}
\end{center}
\caption{Undeformed torus seen from an inclination
of $45^{\circ}$ (left) and $85^{\circ}$ (right). {The color bar, common
to both panels, shows the ratio $\Delta\equiv(\nu_{\mathrm{obs}} / \nu_{\mathrm{em}})^3$,
where $\nu_{\mathrm{obs}}$ and $\nu_{\mathrm{em}}$ are the observed and
emitted frequencies. This is the redshift factor that modulates the observed
intensity}. At high inclination,
the beaming effect {(which makes the redshift factor brighter on the approaching side
of the torus)} is more pronounced and the apparent
area of the secondary image is bigger. }
\label{fig:torusinit}
\end{figure*}

Figure~\ref{fig:torusinit} shows the undeformed torus image
as seen from an inclination of $45^{\circ}$ and $85^{\circ}$. 
The two values of inclination emphasize the bigger impact
of the beaming effect at high inclination, which strongly increases
the dynamic of the image. Moreover, the apparent area of the
primary and secondary images vary, depending on inclination.
The primary image dominates clearly at small inclination, whereas
the secondary image's apparent area becomes important at
high inclination.
The value
of $\beta$ is chosen to ensure that the torus
is indeed slender, i.e., that for all points inside the torus

\be
\label{eq:slender}
\frac{\left|r-r_0\right|}{r_0} \ll 1, \qquad \left|\frac{\pi}{2} - \theta\right| \ll \frac{\pi}{2}.
 \ee

{We choose $\beta=0.05$, a value that satisfies these constraints well
to be able to still abide by the slender torus approximation
even for the deformed (and, in particular, extended) torus.}
This value will be fixed for the remainder of this article. 


%

When the torus is deformed, two main effects will have an impact on
the light curve: the change with time of projected emitting area
and the varying amplitude of relativistic effects as
the torus surface is approaching or receding from the
black hole. To be able to compare the respective 
impacts of these two effects on the various deformed tori, 
we require that:

\begin{itemize}
\item[-] the specific intensity emitted at the surface of the optically
thick torus is inversely proportional to its cross section. This 
will erase the effect of varying the flux by changing the torus
cross-section area for torus expansion (all other deformations leave
the cross-section area unchanged). Only the effect of changing the projected
emitting area will thus remain;
\item[-] the `closest approach' of the torus to the black
hole, as defined by the minimum value of the $r$ coordinate, 
is the same for all deformations. This ensures that the light
emitted by the torus
experiences comparable general relativistic effects, whatever
the deformation. Precisely, we require that at closest approach,
$\bar{x} \approx -2.5$ (compared with the initial minimum 
value of $\bar{x}=-1.5$ for the undeformed torus). We note that
the slender condition in Eq.~(\ref{eq:slender}) is still satisfied.
\end{itemize}

The last condition will also impose the value of the deformation $\lambda$
parameter on all deformations. We stress that $\lambda$ is different
for different deformations to allow a common closest approach
for all deformations.

%
%
\subsection{Light curves and power spectra}
%
%

\subsubsection{Definitions}

Light curves are easily computed by calculating images of
the deformed torus at various times. The time sampling was
chosen to be around $100$ frames per Keplerian period, and
images are $1000\times 1000$ pixels. As the Keplerian
period at radius $r_0=10.8\,M$ is $t_K = 230 \,M$,  
the time interval between two frames is around $\delta t= 2\,M$.
The total number of frames computed for one given light curve
is $N=200$, thus covering around two Keplerian periods.
One point
on a light curve is simply equal to the summation of all pixels
of a given frame (this boils down to summing intensity over all
solid angles and thus to computing a flux).

Power spectral densities are computed in the following way.
Let $F(t_k)$ be the flux value at time $t_k = k\delta t$.
The power spectral density at frequency $f_k=k/ (N \delta t)$
is defined as the square of the modulus
of the fast Fourier transform of the signal, thus:

\be
PSD(f_k)= \left| \sum_{j=0}^{N-1} F(t_k) \mathrm{exp}(2\pi i j k/N)\right|^{2}
\ee

\subsubsection{Observable characteristics of various deformations}

\begin{figure*}[htbp]
\begin{minipage}{1\hsize}
\includegraphics[width=1\hsize]{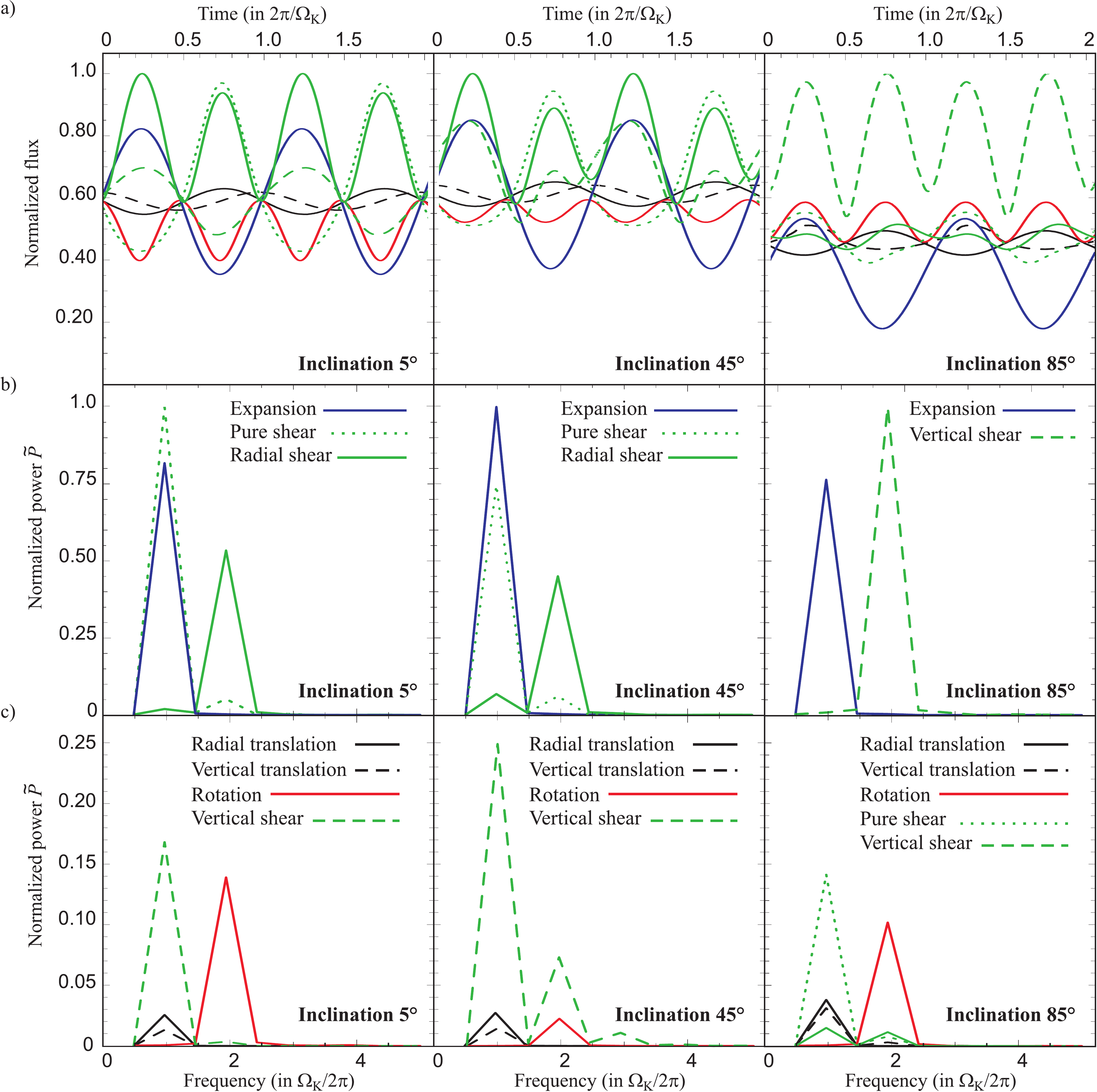}
\end{minipage}
\caption{ a) Normalized light curves for torus deformations. b) Normalized power spectral densities. Set of deformations with large maximal normalized values, $\tilde{P}_{\mathrm{max}}\gtrsim0.5$. b) Normalized power spectral densities. Complementary set of deformations with small maximal normalized values, $\tilde{P}_{\mathrm{max}}\lesssim0.25$. We note that the power spectra are represented as triangular functions for readability only, although they are actually made of a succession of infinitely thin peaks.}
\label{fig:res}
\end{figure*}

{{Figure~\ref{fig:res}a) displays the light curves obtained
for all deformations of the slender torus, seen with 
an inclination parameter of $5^{\circ}$, $45^{\circ}$, and $85^{\circ}$. 
The panels b) and c) of the same figure show the power density spectra of all deformations.}} The spectra 
are presented in two distinct panels for visibility as the power associated with
different deformations or the same deformation at different
inclinations varies significantly.

The biggest modulations (or biggest power densities) are obtained for deformations
that lead to the biggest change of the torus apparent area.
These deformations depend on inclination, but always include
expansion and one or more shears. 

The variation of relativistic effects linked to the changing
torus location is less important than the change of apparent area. 
This is clearly demonstrated by noting
that the smallest modulation is always obtained for vertical translation,
which indeed leads to the smallest change of apparent area.

Whatever the inclination, expansion leads to a high power density,
while translations and rotation give low power densities. On
the other hand, shears exhibit a very inclination-dependent
power density: radial and pure shears give high power for low inclination,
while vertical shears give high power for high inclination. 
All these observable characteristics are directly linked
to the change in apparent area of the deformed torus
at various inclinations.

The time average of the light curves differs from one
deformation to the other. It is dictated by the 
average value of the apparent area.

The dominating frequency of the light 
curves depends strongly on the deformation.
Power is balanced between one time and two times
the Keplerian frequency $\nu_{\mathrm{K}}$.
Translations
and expansion give a single-peak spectrum centered
on $\nu_{\mathrm{K}}$.
Rotation also gives a single-peak spectrum,
but centered on $2\,\nu_{\mathrm{K}}$, 
as all possible values of apparent areas
are covered after only half a Keplerian period in this case only.
Shears give rise to a double-peak structure
in the power spectrum, with a different balance of power
between $\nu_{\mathrm{K}}$ and $2\,\nu_{\mathrm{K}}$
for radial, vertical, and pure shear.

%
%
\subsection{Discussion}
%
%

Figure~\ref{fig:res} shows that the power spectral
density is an interesting probe of the deformation of a slender
accretion torus surrounding a black hole. Different kinds of deformations
lead to significantly different power spectra.

It is interesting to note that the power density associated 
with expansion and the dominant shears is always one
to two orders of magnitude higher than the power density associated
with translations and rotation. This is due to the highly different evolution
of the torus apparent area for these kinds of deformations.

The single- or double-peak nature of the power spectrum is
also an interesting probe of the underlying kind of deformation.
However, the very inclination-dependent aspect of the 
shears power spectra, as opposed to the other kind
of deformations, presents a difficulty.

Taking these two probes into account (maximum power, dominating
frequency) it is possible to constrain the kind of deformation
by studying the power spectrum. Knowing the maximum value of
power density, it is possible to determine whether the underlying deformation
is within \{expansion, shear\} or \{translation, rotation\}. Then,
as a function of the value of the dominating frequency, it is possible
to determine the kind of deformation: expansion, shear, translation, or rotation.
But it is not straightforward to determine what kind of translation
or what kind of shear unless the inclination parameter is already 
constrained.

However, this work is only a first step towards a more sophisticated
treatment of the observable characteristics of non-equilibrium
accretion slender tori. Therefore, it is not the aim of this article to propose
a way of constraining the motion of the accretion torus from the
power density spectrum properties. 





\section{Conclusions and perspectives}
\label{sec:conclu}

We compute light curves and power density spectra of
slender optically thick accretion tori surrounding a Schwarzschild
black hole, which are subject to simple deformations (translations, rotation,
expansion, and shears).
We show that the power spectrum can be used to constrain 
the underlying kind of torus deformation.

{This is a first step towards the full study of the observable
characteristics of oscillating slender accretion tori.
Our conviction is that combining simple deformations
of slender tori will allow realistic oscillations to be mimicked. From this perspective,
our present result that it is possible to differentiate various kinds of simple deformations from
the observed power spectra is promising. Future work will determine
whether this still applies when combinations of simple deformations
are considered and whether this could lead to a practical tool
for studying power spectra of high-frequency QPOs to
constrain the underlying physical model.}
{We believe that such tests can finally bring important outputs in the field of verifying the predictions of general relativity in the exploration of observational data accumulated so far. We also believe that outputs of sophisticated ray-tracing simulations could be key components of the proper interpretation of data received from planned future X-ray missions such as Large Observatory For X-ray Timing \citep[][]{feroci-etal:2012}.}

\begin{acknowledgements}
We acknowledge support from the Polish NCN
UMO-2011/01/B/ST9/05439 grant, the Swedish VR grant, {and the Czech grant GA\v{C}R~209/12/P740. We further acknowledge the project CZ.1.07/2.3.00/20.0071 `Synergy' supporting the international collaboration of IF Opava.}
Computing was partly done using the Division Informatique de
l'Observatoire (DIO) HPC facilities from Observatoire de Paris
(\url{http://dio.obspm.fr/Calcul/}).
\end{acknowledgements}

\bibliographystyle{aa}
\bibliography{OscillatingTorus}

\end{document}